\documentclass[12pt]{iopart}
\usepackage{iopams}  
\usepackage{graphicx}

\begin{document}
\bibliographystyle{iopart-num}

\title{Quasiballistic quantum transport through Ge/Si core/shell nanowires}

\author{D. Kotekar-Patil $^1$\footnote{Present address:Division of Physics and Applied Physics, Nanyang Technological University, Singapore}, B.-M. Nguyen$^2$\footnote{Present address:HRL Laboratories, 3011 Malibu Canyon Road, Malibu, California 90265, USA}, J. Yoo$^2$, S. A. Dayeh$^3$ $^{,4}$ $^{,5}$, S. M. Frolov$^1$}

\address{$^1$ Department of Physics and Astronomy, University of Pittsburgh, Pittsburgh PA, 15260, USA}
\address{$^2$ Center for Integrated Nanotechnologies, Los Alamos National Laboratory, Los Alamos, NM 87545, USA}
\address{$^3$ Department of Electrical and Computer Engineering, University of California, San Diego, La Jolla,CA 92037, USA}
\address{$^4$ Graduate Program of Materials Science and Engineering, University of California, San Diego, La Jolla,CA 92037, USA}
\address{$5$ Department of NanoEngineering, University of California, San Diego, La Jolla, CA 92037, USA}
\ead{dpatil@tnu.edu.sg}
\vspace{10pt}

\begin{abstract}

We study signatures of ballistic quantum transport of holes through  Ge/Si core/shell nanowires at low temperatures. We observe Fabry-P$\acute{e}$rot interference patterns as well as conductance plateaus at integer multiples of 2e$^2$/h  at zero magnetic field. Magnetic field evolution of these plateaus reveals  relatively large effective Land$\acute{e}$ g-factors. Ballistic effects are observed in nanowires with silicon shell thicknesses of 1 - 3 nm, but not in bare germanium wires. These findings inform the future development of spin and topological quantum devices which rely on ballistic subband-resolved transport.

\end{abstract}

%
%
%
%
%

\section{Introduction}

A surge of interest in devices based on nanowires with strong spin-orbit interaction is due to their relevance for quantum computing \cite{Nadj-Perge2010,Pribiag2013,Veldhorst2015, Kotekar-Patil2017,Maurand2016,Heorhii2016,Higgingotham2014} and for the realization of topological superconductivity and Majorana fermions \cite{Mourik2012,Albrecht2016}. 
In this context, spin-orbit interaction in germanium/silicon nanowires was predicted and estimated experimentally to be strong \cite{Kloeffel2011,Hu2007, Hao2010, Brauns2016_2, Azarin2017,Maier2014}, superconducting contacts to these nanowires were demonstrated \cite{Xiang2006,Zu2016}, and large Land$\acute{e}$ g-factors were reported \cite{Brauns2016,Azarin2017}. Moreover, mobility upto 4200 cm$^2$/Vs was reported in these nanowires \cite{Conesa-Boj2017}. These effects, provided they are observed in the ballistic transport regime \cite{Ilse2013,Lu2005,Kretinin2010}, are essential for Majorana experiments. 

In this paper, we report transconductance resonances consistent with one-dimensional (1D) subbands occupied one-by-one as the top gate voltage is made more negative.
 At low temperatures  ($<$1 K) transport is strongly dominated by Fabry-P$\acute{e}$rot interference patterns.
The magnetic field evolution of conductance resonances reveals large g-factors. 
These effects are observed for a range of silicon shell thicknesses without any obvious dependence on shell thickness of silicon shell thickness, however devices without any shell did not show ballistic transport signatures and exhibited substantial charge instabilities. 
We estimate the subband spacing to be $\sim$20 meV, the low temperature mobility of up to 1000 cm$^2$/Vs and the mean free path of 70 nm. The mean free path is larger than the core diameter consistent with the quasiballistic regime.

\section{Fabrication}

Ge/Si core/shell nanowires (NWs) are grown using the low pressure, cold-walled chemical vapor deposition. NWs are grown with various core diameters (15 - 55 nm) and shell thicknesses (0 - 4 nm) \cite{Minh2014}. 
Fig. 1a shows a high resolution transmission electron micrograph of a Ge/Si core/shell NW demonstrating a high degree of control over the silicon shell thickness \cite{Dayeh2011,Dayeh2011_2,Dayeh2013_2}. 
To fabricate devices, the NWs are sonicated in isopropanol and then dropped onto Si$_3$N$_4$ substrates with alignment markers. 
In order to achieve the low ohmic contact resistance, a dip in hydrofluoric acid is performed to etch the native oxide on the silicon shell. 
Electron beam evaporation and electron beam lithography are used to define two 150 nm thick nickel contacts. This step is followed by a 30 second rapid thermal annealing at 300$^\circ$C. 
During annealing, Ni diffuses into the NW from both ends forming highly doped NiGe$_x$/NiSi$_y$ ohmic contacts \cite{Wu2004,Tang2014,Dayeh2013}. 
Segments of the NW between the sections of NiGe$_x$/NiSi$_y$ define the Ge/Si channel length (L = 250 - 450 nm). 
As a last step, a top gate stack consisting of a 10 nm thick hafnium oxide gate dielectric is deposited using atomic layer deposition and then a 30 nm/100 nm thick Ti/Au top gate electrode is evaporated. 
Gate contact overlaps with the NiGe$_x$/NiSi$_y$ region fully covering the unannealed Ge/Si channel.

\begin{figure}
 \centerline{ \includegraphics[width=10cm]{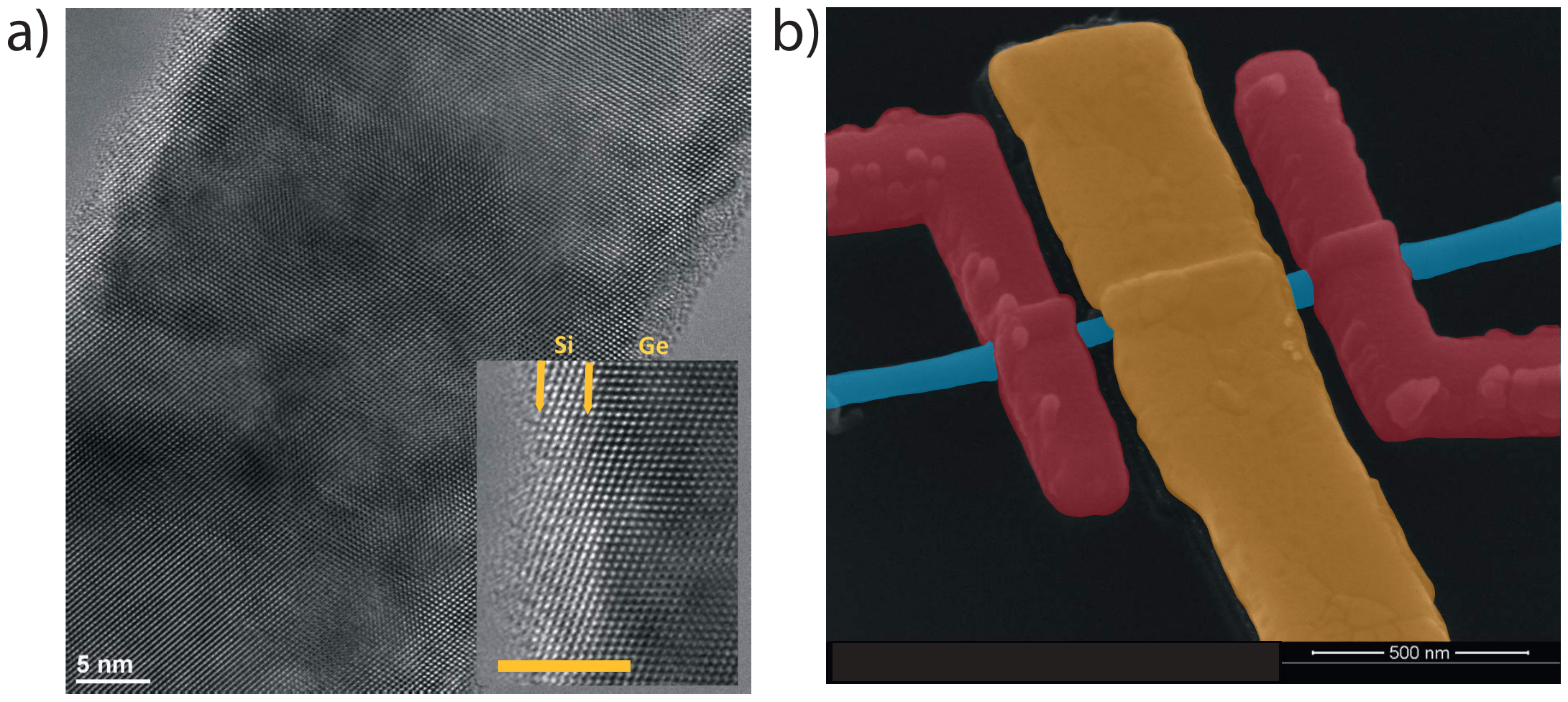}}
\caption{($a$) High resolution transmission electron micrograph of a Ge/Si nanowire. Inset shows a well-defined interface between the Si shell and the Ge core. ($b$) False color scanning electron micrograph of a device where blue region corresponds to the NW, red regions correspond to the source and drain contacts, yellow region corresponds to the top Ti/Au gate.}
  \label{fig.1}
\end{figure}

\section{Singatures of conductance quantization}

Electrical characterization is performed in a dilution refrigerator equipped with a 9 T magnet, using a standard lock-in technique at 27 Hz with an excitation voltage of 50 $\mu$V. 
Noise attenuation is done in 2 stages: at room temperature using $\pi$-filters, and at low temperatures using two-stage low-pass RC filters. 
Room temperature characterization of the same devices is reported by Nguyen et al.\cite{Minh2014}. The room temperature saturation resistance, which is the two-terminal resistance measured at highly negative top gate voltages (V$_g$), is in the range of 10 k$\Omega$.  The room temperature field effect mobility is 150 - 250 cm$^2$/Vs for NWs with silicon shells independent of shell thickness, and 50 cm$^2$/Vs for bare Ge NWs.  Measurements in this work show that at low temperatures the saturation resistance is comparable to the room temperature resistance (1 - 20 k$\Omega$). The low temperature field-effect mobility is in the range of 200 - 500  cm$^2$/Vs with the highest mobility of 1000 cm$^2$/Vs extracted from the pinch-off traces using gate-to-nanowire capacitance calculated by a self-consistent Poisson solver (see supplementary information).

\begin{figure}
\centerline{\includegraphics [width=10cm]{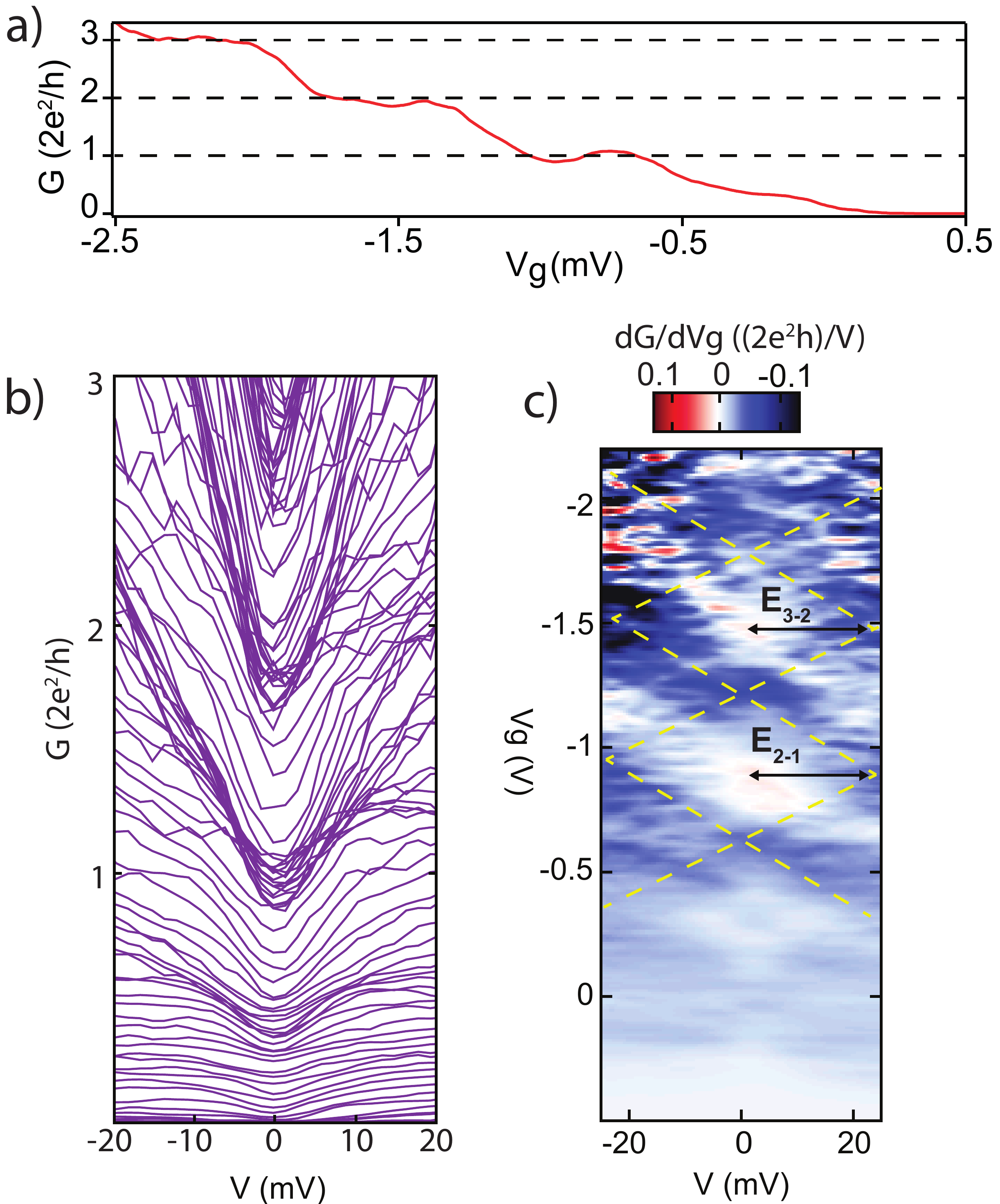} }
\caption{($a$) Conductance G as a function of gate voltage $V_g$, at V = 0 V. ($b$) Waterfall plot of conductance: each  trace is taken at a fixed V$_g$ = 0.5 - (-2.5) V. There is no offset between the traces. ($c$) Transconductance dG/dVg of the data in panel (b) with high transconductance resonances marked by dashed lines. Energy splittings deduced from transconductance resonances are indicated by solid arrows. All data obtained at zero applied magnetic field, T = 5.5. K; a series resistance of 21 k$\Omega$ is subtracted.} 
\label{fig2} 
\end{figure} 

As gate voltage is swept from positive to negative, conductance increases in steps of 2e$^2$/h (Fig. 2a). 
We associate this with one-dimensional spin-degenerate subband-resolved transport. 
Additionally, we observe a conductance plateau below the first 2e$^2$/h plateau. 
Such features are frequently reported in quantum point contacts \cite{Thomas1996,Cronenwett2002,Reilly2002,Komijani2010} and are not the focus of the this work. 
We further investigate this device in the non-linear regime where conductance through the NW is studied as a function of bias voltage (V) and gate voltage. 
Fig. 2b shows the  waterfall plot in which we observe accumulations of conductance traces near 2e$^2$/h and 4e$^2$/h.   
However, plateaus at 0.5 x 2e$^2$/h that are expected in quantum point contacts at high bias, when the bias exceeds the subband spacing, could not be resolved due to strong current fluctuations at high bias.

Fig. 2c shows the transconductance (dG/dV$_g$) of the data in the panel 2b. 
High transconductance resonances move linearly as a function of V$_g$ and V. 
The difference in bias between points where positive and negative slope transconductance resonances meet (forming diamond-shaped regions) indicate the energy separation between the 1D subbands. 
From Fig. 2c, we observe that the first (E$_{2-1}$) and second (E$_{3-2}$) transconductance diamonds have approximately the same size of $\approx$ 22 meV. 
This energy separation is consistent with transverse quantization in the nanowire for heavy holes with an effective mass of $m_{hh}=0.28m_e$, where $m_e$ is the free electron mass.  
Additionally, the slopes of the transconductance resonances are used to extract the gate lever arm parameter $\alpha$ = dV/dV$_g$ = 29.5 meV/V.

\section{Magnetotransport}

\begin{figure}
\centerline{\includegraphics[width=16cm]{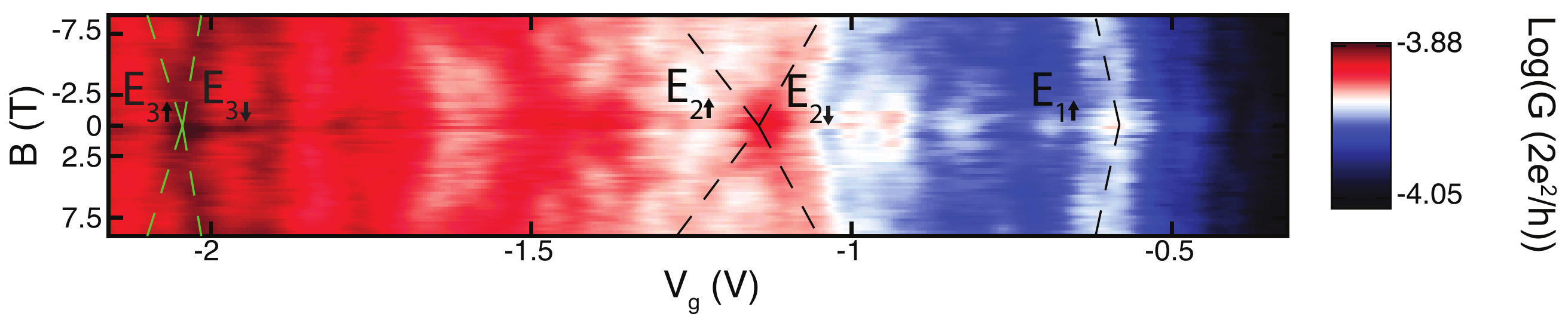} }
\caption{Differential conductance as a function of magnetic field and V$_g$ at V = 0V} for field oriented  at an angle of 45$^\circ$  to the nanowire. Black dashed lines mark the spin splitting of the first ( E$_1$ $_\uparrow$) and second conductance step (E$_2$$_\downarrow$ and E$_2$$_\uparrow$) and green dashed lines mark spin splitting of the third conductance step (E$_3$$_\downarrow$ and E$_3$$_\uparrow$).
\label{g-factor} 
\end{figure}

Fig. 3 shows the evolution of conductance steps as a function of magnetic field. We note that we only observe Zeeman splitting for the second and third conductance steps while the first step only exhibits a Zeeman shift rather than a splitting. In addition, there are other resonances in fig. 3 which move linearly with magnetic field. We associate these resonances to quantum interference effects \cite{Folk2009}, similar g-factors can be extracted for these resonances. 
Zeeman splitting is given by $\Delta$E$_z$ = g $\mu_B$ B, where $\mu_B$ is the Bohr's magneton. 
We use the lever arm parameter calculated from the transconductance diamond in Fig. \ref{fig2}c to convert the V$_g$ axis into the energy scale. 
This gives a g-factor for each transition which we denote by g$_1$=2 for the first transition, g$_2$=10.7 $\pm$ 2.3 for transition between the first and the second conductance steps and g$_3$=4.7 $\pm$ 1.3 for transition between the second and the third conductance step.

Due to the large effective hole masses, the orbital effects of magnetic field are ignored when extracting g-factors, however they may contribute to the shifts of resonances in magnetic field, especially for the resonances with the smaller apparent g-factors.
Large g-factors were also recently observed in Ge/Si nanowire quantum dots \cite{Brauns2016,Azarin2017}.

\section{Fabry-P$\acute{e}$rot interference}

While all devices are fabricated following the same process, of 45 measured devices, only two show signatures of conductance quantization (shown in fig.2 and S1). In a typical device, subband-resolved transport and conductance plateaux are scrambled due to backscattering and are replaced with Fabry-Perot interference within the channel (Fig. 4a).
This device is based on a NW with diameter d = 35 nm, length L = 350 nm and shell thickness t$_{si}$ = 1 nm  at T = 400 mK. 
Mobility extracted from G(V$_g$) trace at  T = 4 K is $\approx$ 450 cm$^2$/Vs.
While this trace does not show conductance plateaus, it is a representative trace for many core/shell devices studied at temperatures below 1 K. 
With the application of negative gate voltage, conductance reaches 4e$^2$/h indicating transport through at least two subbands. 
In the whole V$_g$ range shown in Fig. 4a, conductance trace exhibits quasi-periodic oscillations. 
The smallest period measured  in V$_g$ is approximately $\Delta$ V$_g$ $\approx$ 9 mV. 
Fig. 4b shows the evolution of these quasi-periodic conductance oscillations as a function of V$_g$ and V: the zero-bias features evolve into checkerboard patterns at finite bias. We attribute this to Fabry-P$\acute{e}$rot interference \cite{Biercuk2005,Kretinin2010}. 
The energy spacing of Fabry-P$\acute{e}$rot resonances $\Delta$E = 1 - 2 meV (fig.4c) is linked to the cavity $L_c$ length by: $\Delta$E = $\frac{\hbar ^2 \pi^2}{2 m^* L_c^2}\approx$ 75-110 nm. This implies that the segment of NW over which Fabry-P$\acute{e}$rot interference takes place is  approximately one-third of the NW channel. 

\begin{figure}
\centerline{\includegraphics[width=10cm]{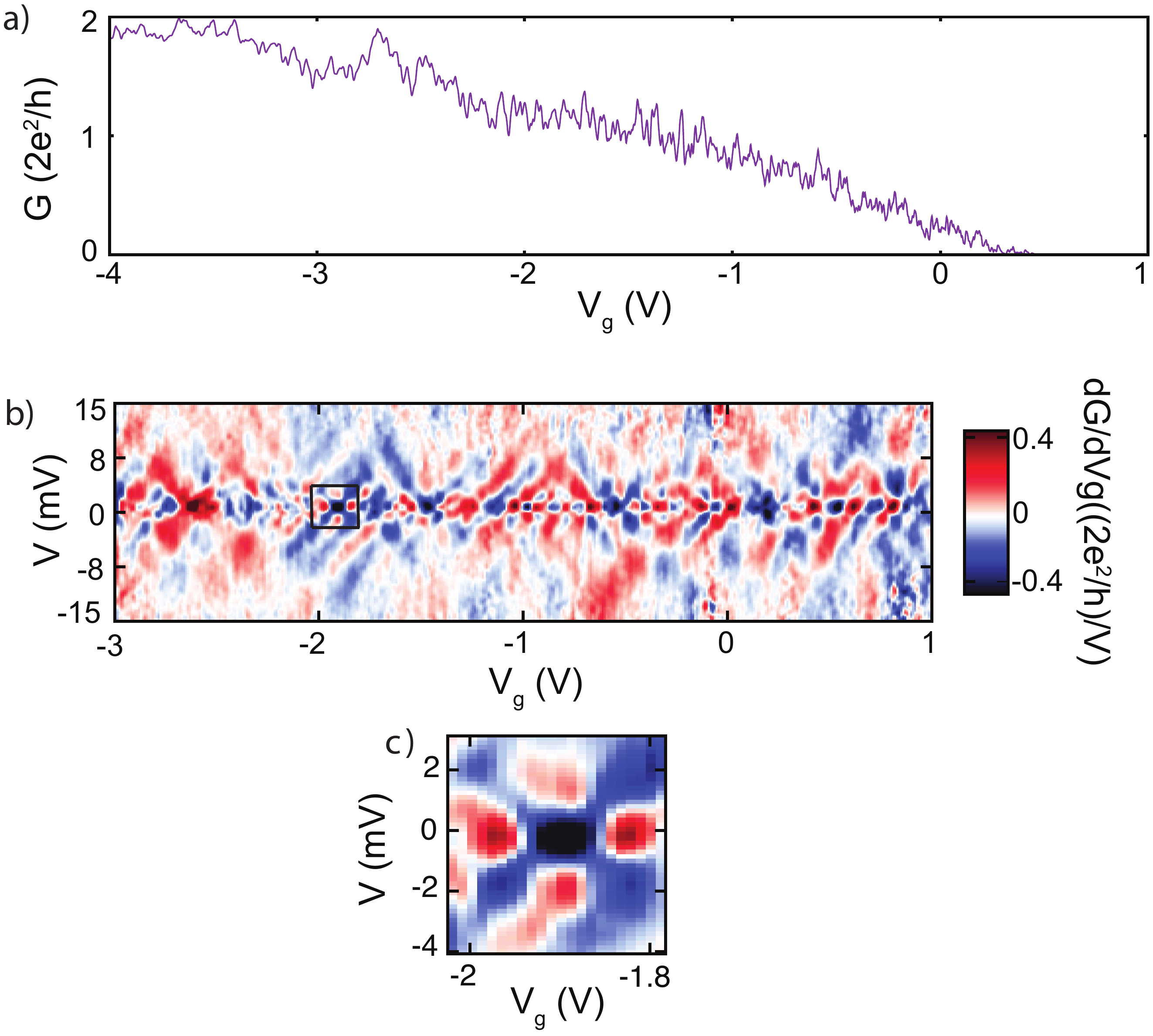} }
\caption{(a) Differential conductance at V = 0 V and (b) transconductance map for a device with silicon shell thickness of 1 nm. T = 400 mK, B = 0 T (c) zoom-in of the checkerboard pattern marked by black box in panel (b).}
\label{fig4} 
\end{figure}

To conclude, signatures compatible with conductance quantization are measured in Ge/Si NW devices. Magnetic field dependence reveals that the hole g-factor in our NWs is large and exhibits strong anisotropy. Moreover, the presence of a silicon shell on the NW results in  quasiballistic transport which is absent in bare Ge NW (see supplementary).

\section{Acknowledgement}

The Ge/Si nanowire growth was performed at the Center for Integrated Nanotechnologies (CINT), U.S. Department of Energy, Office of Basic Energy Sciences User Facility at Los Alamos National Laboratory (Contract DE-AC52-06NA25396) and Sandia National Laboratories (Contract DE-AC04-94AL85000). We thank Z. Su and A. Zarassi for technical help and useful discussions. S.A.D. acknowledges NSF support under DMR-1503595 and ECCS-1351980. S.M.F. acknowledges NSF DMR-125296, ONR N00014-16-1-2270 and Nanoscience Foundation, Grenoble. 

\section{References}
\bibliography{sige.bib}

\end{document}